# Dopant-induced ignition of helium nanoplasmas – a mechanistic study

Andreas Heidenreich[1,2], Dominik Schomas[3] and Marcel Mudrich[4]

[1] *Kimika Fakultatea, Euskal Herriko Unibertsitatea (UPV/EHU) and Donostia International Physics Center (DIPC), P.K. 1072, 20080 Donostia, Spain*

[2] *IKERBASQUE, Basque Foundation for Science, 48011 Bilbao, Spain*

[3] *Physikalisches Institut, Universität Freiburg, 79104 Freiburg, Germany*

[4] *Department of Physics and Astronomy, Aarhus University, 8000 Aarhus C, Denmark*

Email: andreas.heidenreich@ehu.eus

**Abstract.** Helium (He) nanodroplets irradiated by intense near-infrared laser pulses form a nanoplasma by avalanche-like electron impact ionizations even at lower laser intensities where He is not directly field ionized, provided that the droplets contain a few dopant atoms which provide seed electrons for the electron impact ionization avalanche. In this theoretical paper on calcium and xenon doped He droplets we elucidate the mechanism which induces ionization avalanches, termed *ignition*. We find that the partial loss of seed electrons from the activated droplets starkly assists ignition, as the Coulomb barrier for ionization of helium is lowered by the electric field of the dopant cations, and this *deshielding* of the cation charges enhances their electric field. In addition, the dopant ions assist the acceleration of the seed electrons (*slingshot effect*) by the laser field, supporting electron impact ionizations of He and also causing electron loss by catapulting electrons away. The dopants' ability to lower the Coulomb barriers at He as well as the slingshot effect decrease with the spatial expansion of the dopant, causing a dependence of the dopants' ignition capability on the dopant mass. Here, we develop criteria (*impact count functions*) to assess the ignition capability of dopants, based on (i) the spatial overlap of the seed electron cloud with the He atoms and (ii) the overlap of their kinetic energy distribution with the distribution of Coulomb barrier heights at He. The relatively long time delays between the instants of dopant ionization and ignition (*incubation times*) for calcium doped droplets are determined to a large extent by the time it takes to deshield the dopant ions.

## 1. Introduction

Transient nanoplasmas formed from neutral clusters and nanodroplets ($\approx$1-100 nm diameter) by the interaction with intense light pulses ranging from near-infrared (NIR) to hard x-rays are a topic of current research. These studies are motivated by the ability of nanoplasmas to absorb huge amounts of laser energy, leading to extreme states of ionized matter. Related to this, nanoplasmas bear the potential for a number of applications as the generation of energetic electrons and ions [1, 2] including the possibility to induce nuclear reactions in table-top experiments [3, 4], as well as the generation of extreme ultraviolet (XUV) light [5] and attosecond pulses [6, 7].

In the NIR excitation regime, nanoplasma formation is initiated by tunnel ionization (TI) and classical barrier suppression ionization (BSI), and is followed by electron impact ionization (EII) [1, 2]. Whether BSI or EII is the dominating ionization channel, depends on the cluster size and laser pulse parameters. In particular at pulse peak intensities $\leq 10^{16}$ Wcm$^{-2}$ and long pulses (200 fs), avalanche-like EII takes over once TI and BSI have provided a sufficient number of electrons. It has been demonstrated that it suffices to generate a small number of seed electrons in nanoclusters to trigger avalanche EII at laser intensities two orders of magnitude below the threshold intensity of TI of the atomic constituents. This can be achieved either by irradiating the pristine nanoclusters by additional weak XUV pulses [8] or by inserting dopant atoms with low ionization threshold intensity into the host clusters [9-12]. The triggering of avalanche-like ionization we refer to as *ignition*.

Helium (He) nanodroplets are interesting benchmark systems to study the effect of doping on the nanoplasma formation. Due to the high first ionization energy of He of 24.6 eV, the highest of all chemical elements, pristine He droplets are highly inert to NIR radiation. Thus, the effect of doping He nanodroplets with other elements of lower ionization energies is very pronounced. Due to the quantum fluid properties of He nanodroplets, well-defined cluster cores of dopant atoms are formed at predictable positions in the interior or at the droplet surface when doping the droplets with multiple dopant atoms. Besides, the simple electronic structure of the He constituent atoms makes He nanodroplets particularly attractive for model calculations.

In two recent studies [13, 14], we have elucidated the roles of the electronic structure as well as the location of dopants inside (rare-gas) or on the surface (alkali, alkaline earth metals) of the host He droplet in triggering avalanche ionization for relatively long pulses of $t_{\mathrm{FWHM}} = 200$ fs (FWHM of the Gaussian intensity envelope) with peak laser intensities in the range $10^{13} - 10^{15}$ Wcm$^{-2}$. Both the ability of dopant atoms to provide seed electrons as well as their location in close contact with the He droplet were found to be crucial parameters. Counterintuitively, those dopants with the lowest ionization energies (potassium, K) turned out to be the least efficient ones in inducing avalanche ionization of He droplets, whereas high yields of He$^+$ and He$^{2+}$ were measured when doping the droplets with xenon (Xe). This was rationalized by Xe unifying several favorable properties as a seed for the He nanoplasma ignition: Its ability to easily donate more than one seed electron per dopant atom, its location inside the droplet, as well as its low heat of cluster formation which limits shrinkage of the droplets due to evaporation of He atoms during the doping process.

The purpose of our present work is to look into various mechanistic aspects of the ignition process in greater detail. In our previous work we found that ignition does not occur immediately after the initial ionization of the dopant cluster. The time delay between dopant ionization and ignition we termed *incubation time*. It can span an appreciable part of the laser pulse duration for calcium (Ca) and K dopant clusters, whereas incubation times are much shorter for Xe dopants. Thus, the question arises which factors determine the length of the incubation time and actually determine the occurrence of ignition. Related to it is the question of the dopants' role in the ignition process at the microscopic level and what determines the dopants' capability of inducing ignition. In this paper we shall restrict our discussion to Ca and Xe as examples for dopants with long and short incubation times, respectively. We shall consider pulse peak intensities $I_M = 10^{13} - 10^{14}$ Wcm$^{-2}$ which is the interesting range for dopant-induced ignition for 200 fs pulses; note that intensities $I_M \gtrsim 5\times 10^{14}$ Wcm$^{-2}$ are needed for igniting He by TI in the absence of dopant atoms.

## 2. Simulations

We previously described the molecular dynamics (MD) simulation method for the interaction of a cluster with a linearly polarized NIR Gaussian laser pulse [13,15,16]. Starting with a cluster of neutral atoms, electrons enter the simulation when the criteria for TI, BSI or EII apply. This is inquired at every MD time step, calculating the local electric field at the atoms consisting of the external laser electric field and the contributions of all ions and electrons of the cluster. Instantaneous TI probabilities are obtained by the Ammosov-Delone-Krainov formula [17], EII cross sections by the Lotz formula [18] with the ionization energy as the atomic Coulomb barrier in the cluster [19]. The effect of chemical bonding on the valence shell ionization energies of Ca dopants is neglected. The trajectories of atoms and electrons are propagated classically.

Interactions between ions are described by Coulomb potentials, ion-electron and electron-electron interactions by smoothed Coulomb potentials. Interactions involving neutral atoms are disregarded except for a Pauli repulsive potential of 1.1 eV [20] between electrons and neutral He atoms. The binding potentials of $He_2^+$ and of other $He_n^+$ complexes are not implemented, so that the $He_2^+$ formation cannot be accounted for.

Three-body electron-ion recombination is automatically accounted for in the MD simulations but does not play an important role at the early stages of the nanoplasma formation considered in this work. The ion charges given in this paper are throughout bare ion charges without electron-ion recombination. The intensity maxima of the Gaussian pulses are in the range $I_M = 10^{13} - 10^{14}$ Wcm$^{-2}$, their temporal width is $t_{FWHM} = 200$ fs (FWHM of the Gaussian electric field envelope $\tau = 283$ fs) which is the pulse width in our previous work. Since in this paper we are interested in the dynamics of doped droplets at distinct pulse peak intensities rather than in a comparison with experiments, MD results are not focally averaged, unlike in our preceding work [13, 14]. However, whenever indicated in the text, MD results are averaged over a set of 50 or 100 trajectories with the same dopant size and pulse parameters but different initial conditions (slightly different initial atomic coordinates and a different seed for the random number generator for TI).

For the He droplets, we use a fcc structure with an interatomic distance of 3.6 Å [21]. We choose He$_{2171}$ as a sample of the droplet size distribution under experimental conditions [13, 14]. The dopant clusters are densest packed assemblies of tetrahedrons and form, as far as possible, spherical shapes. The interatomic distances are: Ca-Ca 3.9 Å (average value for Ca clusters [22], Xe-Xe 4.33 Å (bulk), He-Xe 4.15 Å [23],

He-Ca 5.9 Å [24]. In case of surface doping with Ca, we assume a dimple depth of 7 Å [25]. The notation for doping the droplet interior will be C, and X and Y for surface doping in parallel and perpendicular orientation of the dopant-He$_N$ complex with respect to the laser polarization axis, respectively.

## 3. Results and discussion

*3.1 Deshielding of the dopant cations*

In this section we describe the microscopic processes which take place in a doped He nanodroplet at the rising edge of the NIR laser pulse. This links to our previous work where the interplay between dopant and He ionization during nanoplasma formation was studied [13, 14]. We shall show that a partial loss of seed electrons plays an important role in the opening of the EII channel of He in Ca doped droplets. Figure 1 exhibits a single-trajectory example of the time-dependent ionization of a He$_{2171}$ droplet doped with a Ca$_{15}$ cluster at the droplet surface, in the low pulse peak intensity regime, $I_\mathrm{M} = 10^{13}$ Wcm$^{-2}$. Shown are several characteristic time-dependent quantities: (a) The normalized electric field envelope of the laser pulse, (b) the number of ionizations and nanoplasma electrons, (c) the ionization energies of He atoms modified by the electric field of the laser and of the ions and electrons (*inner field*) as well as the average kinetic energy of the electrons, (d) functions characterizing the propensity for EII at He, and (e) electron populations exceeding indicated kinetic energies.

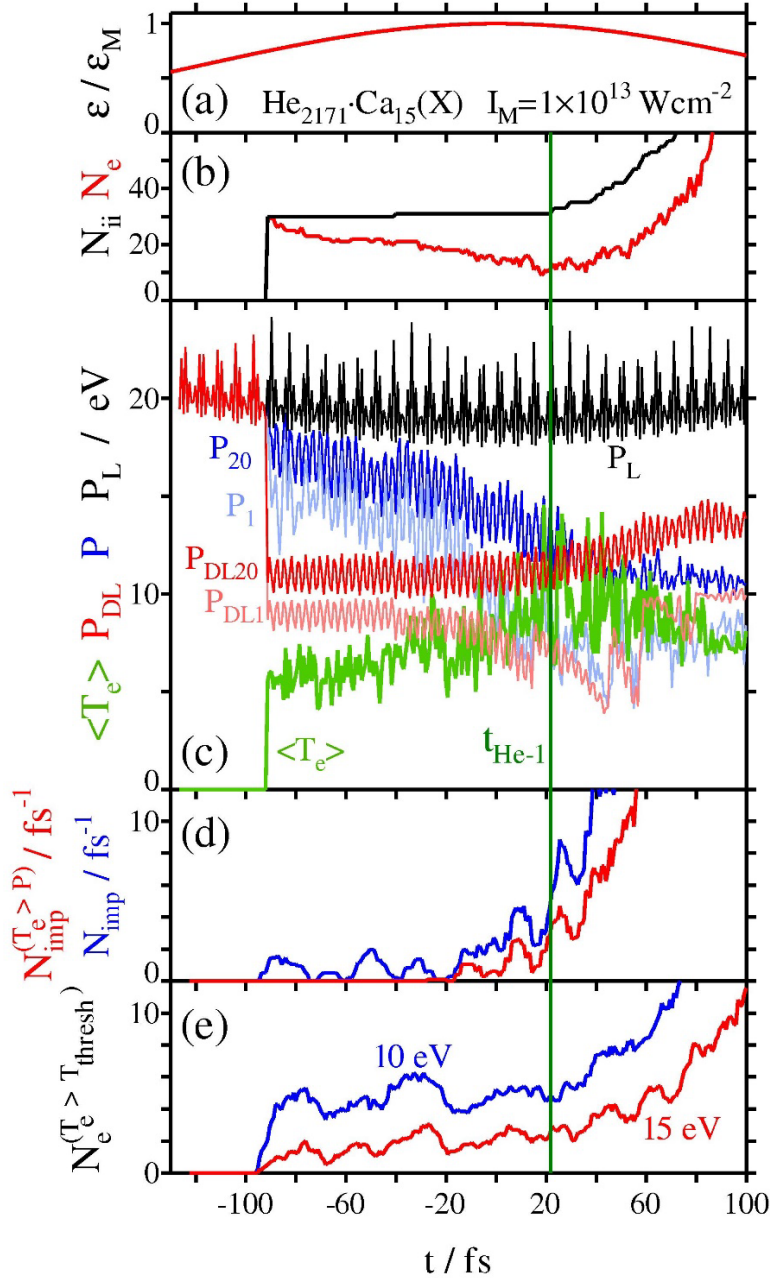

Figure 1: A single-trajectory example of the early stage of the ionization dynamics of a He$_{2171}$ droplet doped with a Ca$_{15}$ cluster. The Ca$_{15}$ dopant is located on the droplet surface such that the dopant-droplet axis is parallel to the laser polarization axis (x direction). The pulse peak intensity is $I_M = 10^{13}$ Wcm$^{-2}$. (a) Normalized electric field envelope. (b) The total number $N_{ii}$ of inner ionizations (black) of Ca and He and the number of nanoplasma electrons $N_e$ (red) inside the droplet nuclear framework and its vicinity (1.5 droplet radii; radii are taken as the distance of the outermost He atom from the center-of-mass of the He host droplet). (c) Ionization energies P of neutral He atoms, i. e., the height of the Coulomb barriers in the time-dependent laser and inner electric field of ions and electrons. The local Coulomb barriers of He atoms form a time-dependent distribution; $P_1$ and $P_{20}$ (light and dark blue curves) represent the first and the twentieth-lowest samples of the distribution. To demonstrate the effect of the inner field, the ionization energies $P_L$ (black curve) under the influence of the laser field alone, and under the sum of laser and dopant ions without nanoplasma electrons, $P_{DL1}$ and $P_{DL20}$ (light and dark red curves are included). The average nanoplasma electron kinetic energy $\langle T_e \rangle$ (green) is added as a measure for the overlap of the $P$ and $T_e$ distributions. (d) The impact count functions $N_{imp}$ representing the number of penetrations of He spheres (radius 1.8 Å) per femtosecond and the number of penetrations $N_{imp}^{(T_e>P)}$ which also meet the energy criterion for EII. (e) The number $N_e^{(T_e>T_{thresh})}$ of nanoplasma electrons with kinetic energies exceeding 10 (blue) and 15 eV (red). The time $t_{He-1}$ of the first He ionization is marked by a vertical green line in panels (b-e). The zero point of the time axis refers to the instant of the maximum of the Gaussian laser pulse envelope.

The process begins with TI and BSI of the valence shell of the Ca atoms at the rising edge of the laser pulse at $t = -91$ fs (figure 1(b)). The dopant ionization starts with a single TI or BSI of one of the dopant atoms, and the rest of the valence ionizations of the dopant cluster follows almost instantaneously, as the inner field in the dopant cluster is enhanced by the first ionization. In this way, the number $N_{ii}$ of inner ionizations (subsuming the generation of nanoplasma electrons by TI, BSI and EII, as opposed to *outer ionization* which denotes the removal of nanoplasma electrons from the droplet nuclear framework and its vicinity [26]) and the number $N_e$ of nanoplasma electrons within or close to the nuclear framework (1.5 cluster radii) rises to 30 (two per Ca atom). At $t = 22$ fs, the ionization of the first He atom occurs (marked by a green vertical bar in figure. 1). Subsequently, at $t \approx 40$ fs, $N_{ii}$ rises steeply, constituting the onset of avalanche-like EII of the He droplet. As described in our previous papers [13, 14], He ionization is always induced by EII; the role of the dopant is to provide the seed electrons and to assist EII by lowering the Coulomb barrier at He by the field of the dopant cations.

During the incubation time, EII of He competes with a partial loss of the seed electrons by outer ionization which reduces the chances for ignition. We found [13, 14] that, depending on slight variations of the trajectories' initial conditions but for the same pulse parameters, either He ionization does not occur at all, terminates after a few He atoms, or ignition takes place. Likewise, the instant of ignition during the laser pulse is highly dependent on the initial conditions. Consequently, for a quantitative comparison with experiment (conducted in our previous papers [13, 14]), one has to average over multiple trajectories. The ignition probability is then given by the fraction of the trajectories where ignition takes place. We found that the ignition probability increases with the dopant size and the laser pulse peak intensity.

To discriminate trajectories with just a few He ionizations, we previously defined a threshold of an average He charge $\langle q_{He} \rangle = 0.1$ as an empirical criterion for detecting ignition [13, 14]. We found particularly long incubation times for dopants of low ionization energies ($K_n$, $Ca_n$) and short ones for $Xe_n$ dopants.

Obviously, since ignition is induced by EII, a necessary but not sufficient condition for its occurrence is an overlap of the electron kinetic energy distribution with the distribution of He ionization energies. A snapshot of the time-dependent distribution of He ionization energies is depicted in figure 2. The close-up view in the inset shows the overlap of the two distributions. Both the laser electric field and the droplet's inner field created by cations and nanoplasma electrons lower the ionization energy (24.6 eV) of isolated He atoms; the resulting Coulomb barriers at a distance of a few Å from the He nuclei determine the new ionization energies in the droplet. Thus, the ionization energies oscillate with the laser frequency. The effect of the inner field on the He ionization energies is particularly large at the He-dopant interface which consists of a few tens of He atoms, making up the low-energy tail of the $P_{He}$ distribution. The effect is quite strong and can lower the He ionization energies by 10-15 eV. Figure 1(c) shows the lowest and the 20$^{th}$ lowest ionization energy of the distribution, $P_1$ (light blue) and $P_{20}$ (dark blue), as a function of time. That is to say, $P_1$ and $P_{20}$ approximately mark the range of ionization energies of the He atoms in the droplet-dopant interface. To monitor the contributions of the laser and of the inner field to the corrected He ionization energies, also the He ionization energies modified only by the laser ($P_L$, black curve, all He ionization energies lowered by the same amount) as well as by the bare dopant cations and the laser electric field ($P_{DL1}$ and $P_{DL20}$, light and dark red curve, respectively), are shown in figure 1(c). $P_L$, $P_{DL1}$ and $P_{DL20}$ were subsequently calculated for the particle coordinates of the original trajectory, that is to say, without altering the course of the trajectory and number of inner ionizations.

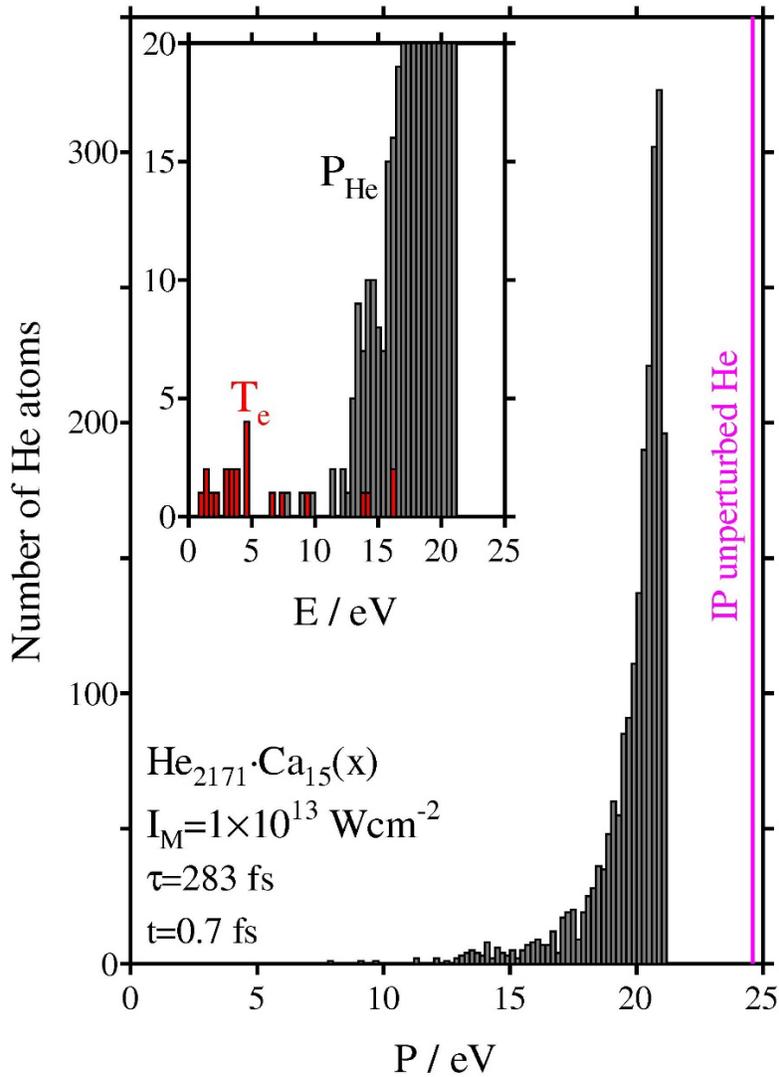

Figure 2: A snapshot of the time-dependent ionization energy distribution of neutral He atoms at $t = 0.7$ fs. Inset: The magnified $T_e$ and the low-energy tail of the $P$ distribution.

At $t = -91$ fs, the ionization energies of the He atoms in the droplet-dopant interface drop significantly below the black $P_L$ curve along with the ionization of the Ca atoms. The red $P_{DL}$ curves drop even more, since they disregard the contribution of the nanoplasma electrons which partially shield the dopant cations (figure 1(c)). Until the instant of ionization of the first He atom at $t_{He-1} = 22$ fs (marked by the green vertical line), the $P_1$ and $P_{20}$ curves drop almost to the level of the corresponding $P_{DL1}$ and $P_{DL20}$ curves, as part of the nanoplasma electrons are removed from the droplet and its close vicinity, as indicated by the number of nanoplasma electrons in figure 1(b) (red curve). Around $t_{He-1}$, the energy range $P_1 \leq P \leq P_{20}$ drops down to the level of the average electron kinetic energy $\langle T_e \rangle$ (green curve in figure 1(c)), suggesting that more seed electrons are energetically able to cause EII of He.

To further analyze the opening of the EII channel of He and the decisive role of the lowering of the He ionization energies, we monitor the trajectory in three different ways:

(i) Spatial overlap of the seed electron cloud with the He droplet. Each He atom is represented by a sphere with a radius of 1.8 Å (half of the He-He distance in the unexpanded droplet before irradiation with the laser pulse) and the number $N_{imp}$ of times electrons enter He spheres is counted, blue curve in figure 1(d).

(ii) Spatial and energetic overlap. We count the number $N_{imp}^{(T_e>P)}$ of such events where electrons enter He spheres with kinetic energies exceeding the ionization energy of the corresponding He atom, red curve in figure 1(d).

(iii) The number $N_e^{(T_e>T_{thresh})}$ of electrons with kinetic energies $T_e$ exceeding threshold values of $T_{thresh} = $ 10 and 15 eV, figure 1(e). These threshold values reflect roughly the ionization energies of He atoms in the dopant-droplet interface region, cf. figure 1(c).

The impact function $N_{imp}$ (blue curve in panel (d)) which accounts only for the spatial overlap of the seed electron trajectories with the He atoms, assumes values $> 0$ from the instant of the Ca ionization on. In contrast, the impact function $N_{imp}^{(T_e>P)}$ (red curve in panel (d)) which accounts for both spatial and energetic overlap assumes values $> 0$ only with a considerable delay, from $t \approx -20$ fs on, thus showing the feature of an incubation time. The electron populations $N_e^{(T_e>T_{thresh})}$ above 10 and 15 eV, panel (e), do not change significantly from the instant of the Ca ionization to the instant of the He ionizations. That is to say, since the instant of the dopant ionization electrons in the kinetic energy range $> 10$ eV are available to induce EII, once the ionization energies at He are lowered sufficiently. Note that the electron populations $N_e^{(T_e>T_{thresh})}$ in panel (e) are a fraction ($\leq 6$) of the entire nanoplasma electron population $N_e$ (20-30) (panel b) until the occurrence of the first He ionization.

In summary, we find that EII of He is triggered by the He ionization energies falling below the kinetic energies of a substantial number of seed electrons (10-15 eV in this example). The lowering of ionization energies of He is caused by a partial loss of seed electrons in the dopant-droplet interface region. The loss of seed electrons progressively deshields the dopant cation charges, so that the cation electric field eventually lowers the Coulomb barriers at He almost to the level caused by the bare cation charges.

Figure 1(c) exhibits the effect of the electric field of the dopant cations and seed electrons on the ionization energies of He. That is to say, the ionization energies of He were recalculated along an already existing trajectory without changing its original course and number of inner ionizations. It is, however, also instructive to demonstrate the effect of the inner field on the ignition itself. To this end, we carried out comparative MD simulations in which part or all inner field contributions were switched off, thus allowing the modified ionization energies to alter the number of inner ionizations during the propagation of the trajectories. Figure 3 shows three examples for the He$_{2171}$ droplet doped with a Ca$_{15}$ dopant cluster at various pulse peak intensities. Shown are histograms of ignition times (based on the $\langle q_{He} \rangle = 0.1$ threshold criterion) together with ignition probabilities obtained from sets of 100 trajectories. The ignition times of the standard MD simulations, in which the ionization energies of He contain all contributions (i. e., the contributions of dopant cations D + He cations H + nanoplasma electrons E + laser L, corresponding to ionization energies $P$ in figure 1(c)), are presented in red. MD simulation results where the He ionization energies were corrected only by the laser electric field (corresponding to ionization energies $P_L$ in figure 1(c)) are represented in blue and results where the effect of the nanoplasma electrons on the He ionization energies was switched off (i. e., contributions D + H + L, ionization energies P$_{DHL}$, not shown in figure 1(c)) in green. Clearly, the effect of the inner field on ignition times and probabilities is large: The ignition probabilities $p_{ignit}$ decrease in the order of the ionization energies $p_{ignit}^{(DHL)} > p_{ignit}^{(DHEL)} > p_{ignit}^{(L)}$ and the instants of ignition are shifted to later times. With increasing pulse peak intensity, ignition is shifted to earlier times and the ignition probability increases. For $I_M = 1 \times 10^{13}$ Wcm$^{-2}$ the ignition probability is zero for trajectories with ionization energies $P_L$, that is to say, inner field assistance is essential to induce ignition. When the shielding effect of the nanoplasma electrons is switched off (ionization energies $P_{DHL}$, green bars), BSI replaces EII as the dominating ionization channel.

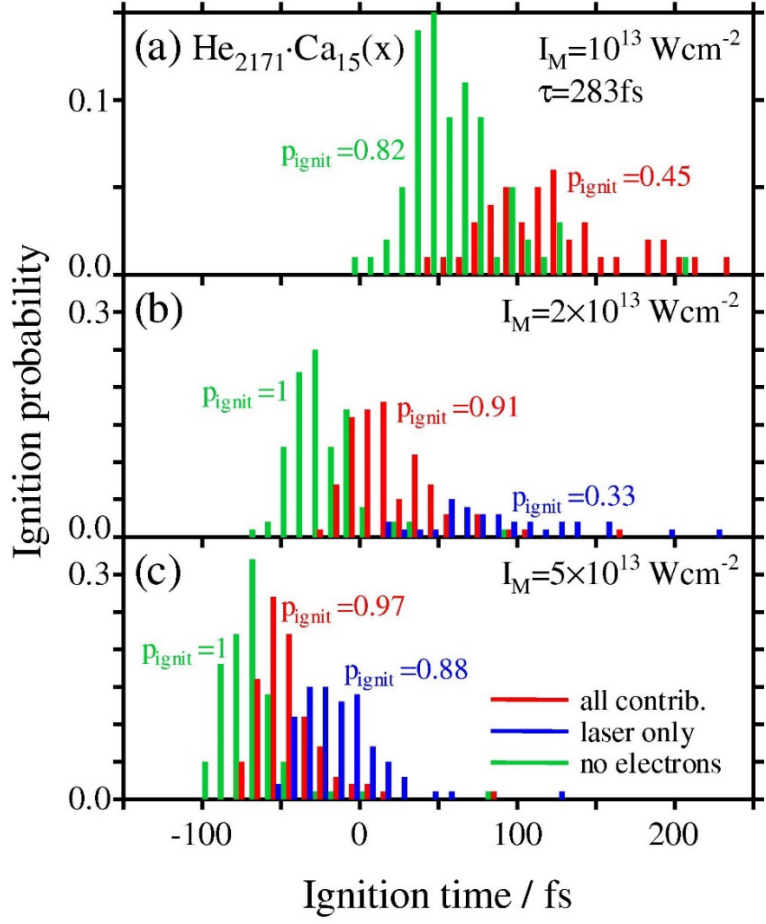

Figure 3: Three examples for the effect of the inner electric field on the ignition time distribution. Red histograms: Standard MD simulations where the Coulomb barriers at He are calculated from the laser and all inner field contributions (ions and electrons). Blue histograms: Coulomb barriers at He without inner field. Green histograms: Coulomb barriers at He calculated only from the laser field, He and dopant cations contributions without the shielding effect of the nanoplasma electrons. The ignition probabilities $p_{ignit}$ are given in each panel.

*3.2 Quasi-stochastic electron dynamics and the slingshot effect of the dopant cations*

In this section we focus on the effect of the seed electron dynamics on the ignition propensity. Figure 4 shows a comparison of two trajectories for $He_{2171} \cdot Ca_{15}$ at $I_M = 1 \times 10^{13}$ Wcm$^{-2}$ with different initial conditions (slightly different initial atomic coordinates and a different seed for the random number generator for TI). In panel (a) the total number $N_{ii}$ of inner ionizations and nanoplasma electrons $N_e$ in or close to the doped droplet is depicted. Trajectory 1 (represented by the blue curves and already discussed in figure 1) leads to ignition, whereas trajectory 2 (red curves) exhibits only two He ionizations. Panel (b) shows the He ionization energies $P_1$ (dotted lines) and $P_{20}$ (closed lines) of both trajectories as well as in grey the $P_{20}$ functions of the rest of the trajectory set. To avoid a congested diagram, the oscillations of $P_1$ and $P_{20}$ caused by the laser electric field are omitted in the presentation (but not in the MD simulation).

Until the first He ionization of trajectory 1 occurs, the $P_{20}$ functions of the entire trajectory set show a considerable spread of up to 3.2 eV which certainly contributes substantially to the different ignition propensities among the trajectories of the trajectory set. However, the differences of $P_1$ and $P_{20}$ between trajectories 1 and 2 are minor; nevertheless ignition takes place only in trajectory 1. This suggests that there is another reason for the occurrence of ignition.

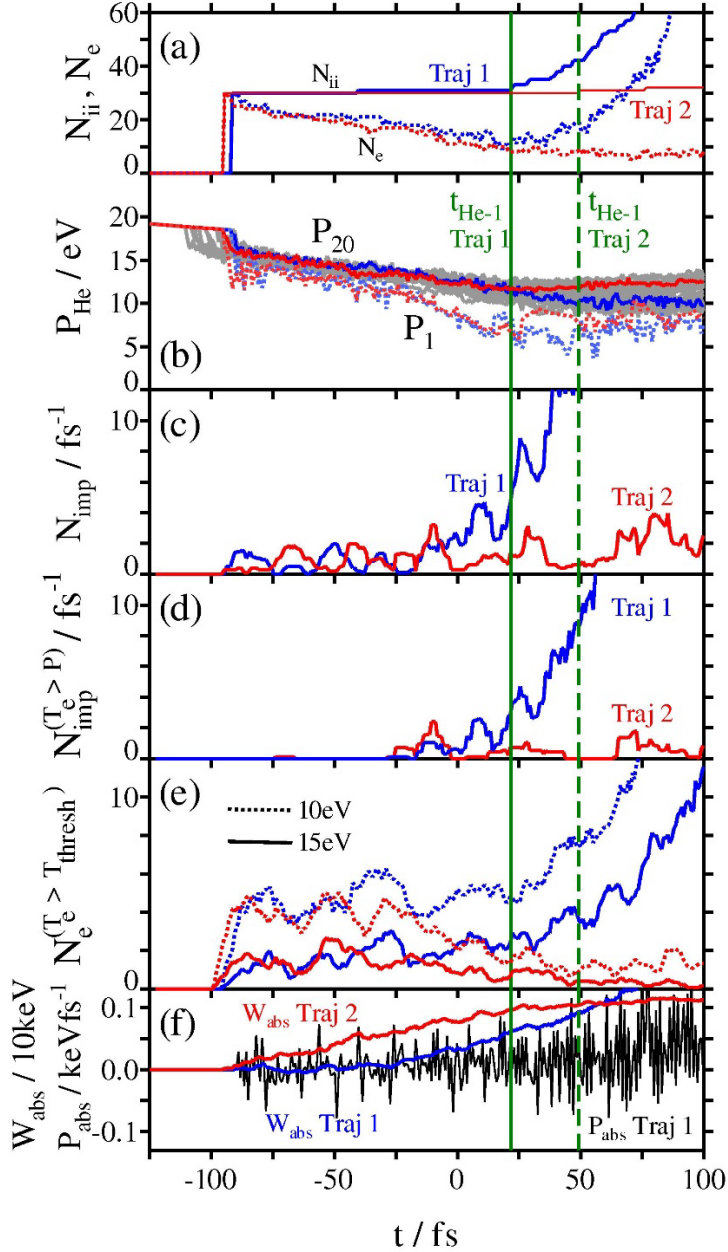

Figure 4: Comparison of the time evolution of two trajectories for $He_{2171} \cdot Ca_{15}(x)$ at $I_M = 1 \times 10^{13}$ Wcm$^{-2}$ with different initial conditions. Throughout the figure, the blue curves represent results for trajectory 1, the red curves for trajectory 2. Trajectory 1 was already discussed in figure 1. (a) The number $N_{ii}$ of inner ionizations (closed lines) and the number $N_e$ of nanoplasma electrons within 1.5 droplet radii (dotted

lines). (b) The ionization energies $P_1$ (dotted lines) and $P_{20}$ (closed lines) of the ionization energy distribution of neutral He atoms. The $P_{20}$ functions of the rest of the set of 100 trajectories are included in grey to demonstrate the spread of $P_{20}$. (c) The $N_{\text{imp}}$ impact count functions. (d) The impact count functions $N_{\text{imp}}^{(T_e>P)}$. (e) The number $N_e^{(T_e>T_{\text{thresh}})}$ of electrons whose kinetic energies exceed 10 eV (dotted lines) and 15 eV (closed lines). (f) The energy absorption $W_{\text{abs}}$ of the electrons. The instantaneous power absorption $P_{\text{abs}}$ (black) is included only for trajectory 1. The times $t_{\text{He-1}}$ of the first He ionization is marked by vertical green lines.

Differences between trajectories 1 and 2 become apparent when looking at the $N_{\text{imp}}$ (figure 4(c)) and $N_{\text{imp}}^{(T_e>P)}$ functions (figure 4(d)). Unlike $P_1$ and $P_{20}$, $N_{imp}$ and $N_{imp}^{(T_e>P)}$ of trajectory 1 assume considerably larger values during the last 25 fs before the first He atom is ionized in trajectory 1, indicating that the spatial as well as the energetic overlap is more favorable for He ionization. Since the energetic part of the $N_{imp}^{(T_e>P)}$ function depends on both ionization energies and electron kinetic energies, it is *per se* not obvious which of the two energy quantities is decisive. However, in view of the similar ionization energies in this pair of trajectories one may suspect that the electron kinetic energy makes the major difference. To elucidate this point further, in figure 4(e) we depict the populations $N_e^{(T_e>T_{\text{thresh}})}$ of nanoplasma electrons exceeding the energy thresholds $T_{\text{thresh}}$ of 10 and 15 eV (as shown in panel (b), this is the energy range required to induce EII). Indeed, trajectory 1 (blue curves) generates considerably more energetic electrons at the onset of He ionization despite of a somewhat higher total seed electron population $N_e$ (panel (a)). The question is then what leads to the higher electron kinetic energies.

To illustrate the electron dynamics in the system prior to ignition, figure 5 exhibits an example of the kinetic energy (blue curve, left ordinate) and distance from the dopant center-of-mass (COM, red curve, right ordinate) of a single, arbitrarily chosen electron of a single trajectory as a function of time. Kinetic energy and COM distance show oscillations with random amplitudes and overall irregular and quasi-stochastic time dependence. For small COM distances, the electron can reach 10-15 eV of kinetic energy. At $t = 15$ fs, even 45 eV are reached whereupon the electron is catapulted to high distances from the dopant COM beyond 100 Å where its kinetic energy then drops to small values. The attained energies are much higher than the maximum kinetic energy of a free electron during a laser cycle which is 1.4 eV at the pulse peak ($I_M = 1 \times 10^{13}$ Wcm$^{-2}$). Apparently, the dopant cluster in combination with the laser field acts as a slingshot for the seed electrons. The laser energy $W_{\text{abs}}$ absorbed by the nanoplasma electrons is the time integral of the power absorption $P_{\text{abs}}$,

$$W_{\text{abs}}(t)=\int_{-\infty}^{t} P_{\text{abs}}(t')dt', \text{ where } P_{\text{abs}}(t)=-\sum_i(e\vec{v}_i\cdot\vec{\varepsilon}), \qquad (1)$$

$e$, $\vec{v}_i$ and $\vec{\varepsilon}$ are the elementary charge, the velocity vector of electron i and the laser electric field, respectively. Since the essential part of the power absorption is the scalar product $\vec{v}_i\cdot\vec{\varepsilon}$ of the velocity of electron *i* and the laser electric field vector, the power absorption will be considerably enhanced, if the electron has already gained velocity from an acceleration in the cation potential. Thus, close fly-bys can confer high kinetic energies to the electron, if the laser electric field has the correct phase. To characterize the efficient acceleration mechanism we adopt the term *slingshot effect* because of its pithiness, although the better physical analog is *powered flyby* or *Oberth effect* in astronautics. The intermittently high electron kinetic energies acquired in this way can induce EII at He or can catapult electrons into high orbits. In the latter case, the electrons are lost as EII seeds but their removal from the vicinity of the dopant cluster also reduces the shielding of the dopant cation charges. There are two other mechanisms which can cause a seed electron to be lost during the incubation time: scattering by the Pauli repulsive potential at the neutral He atoms and electron-electron repulsion. We have not assessed the relative importance of the three loss mechanisms.

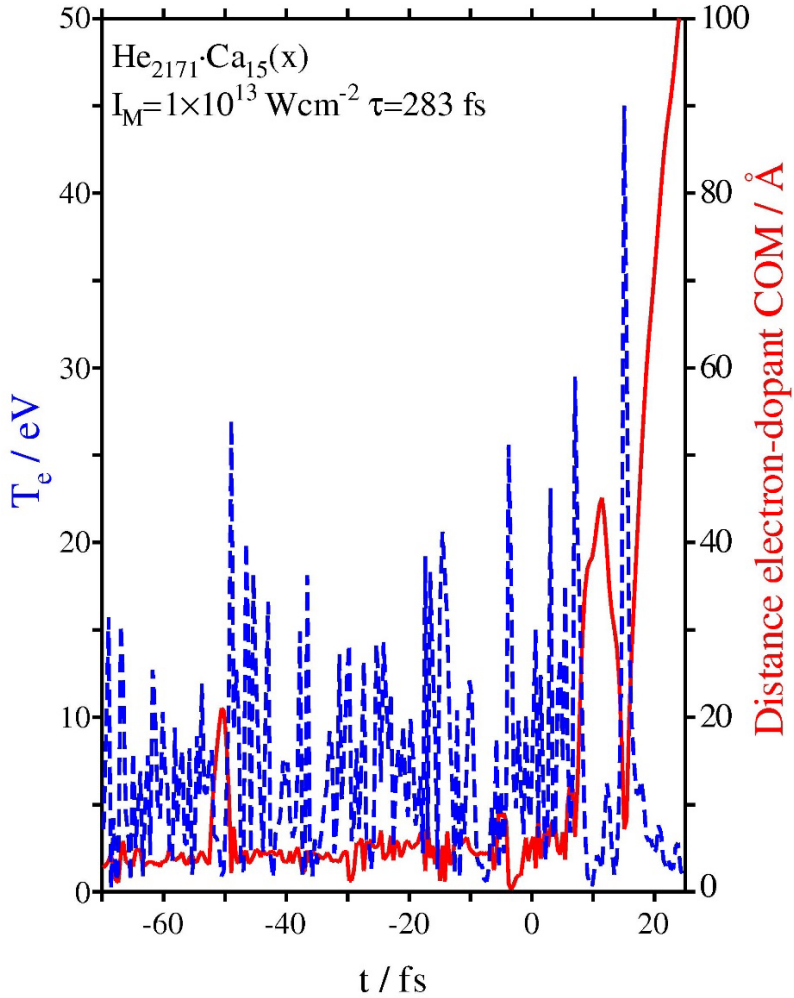

Figure 5: Illustration of the quasi-stochastic electron dynamics: Kinetic energy (blue curve, left ordinate and distance from the dopant center-of-mass (COM, red curve, right ordinate) as a function of time for a single, arbitrarily chosen electron of a single trajectory.

The energy absorption $W_{abs}(t)$ is positive during the incubation time, as shown for trajectories 1 and 2 in figure 4(f). For trajectory 1, $W_{abs}(t)$ increases steeply during the avalanche ionization $t \gtrsim 0$ (only the onset of the steep increase is shown), while $W_{abs}(t)$ converges to 1.2 keV for trajectory 2 where the avalanche does not evolve. In our previous work [14] we discussed the occurrence of a nanoplasma resonance of the seed electrons already during the incubation time. Although a positive $W_{abs}(t)$ as a necessary condition for resonance is met, the presence of a resonance seems to be ambiguous in the present case, since the scalar products $\vec{v}_i \cdot \vec{\varepsilon}$ are positive only during short times (in many cases ½ - 1 laser cycle) of the very irregular electron paths.

In conclusion, the seed electrons can temporarily acquire high kinetic energies (typically for fractions of a laser cycle) by dopant-assisted laser energy absorption (*slingshot effect*). This can provide sufficient kinetic energy for EII but also for catapulting seed electrons away from the doped droplet. The dynamics of the seed electrons is very irregular and quasi stochastic. The reason for ignition not occurring in one of the two discussed trajectory examples is differences between the populations of energetic seed electrons. That two trajectories with only slightly different initial conditions lead to a completely different outcome, manifests the chaotic behavior of the highly non-linear dynamics in this classical many-body system.

*3.3 Dopant mass effect*

If the picture of the dopant cation cluster acting as a slingshot is correct, its accelerating effect on electrons should decrease as the dopant spatially expands, since the potential well created by the dopant cations becomes shallower. To elucidate the effect of the dopant expansion, we have carried out comparative MD simulations with fictitious Ca masses, for 4 (He) and 131 (Xe), as well as for 40 (real Ca mass) atomic mass units (amu). In these simulations, He ionization was disabled in order to test the dopant's effect alone without the feedback of He ionization on the nanoplasma electron population. Figure 6 shows trajectory set-averaged results. Panel (a) shows the relative radius $R/R_0$ of the expanding dopant cluster. Panel (b) exhibits the electron populations at kinetic energies >10 and >15 eV which decrease indeed fastest for $^4$Ca and most slowly for $^{131}$Ca. The rates at which the $P_{20}$ functions evolve in time, obey the same mass dependence, that is, with increasing dopant mass the $P_{20}$ minimum is passed at a later time (panel (c)). The electron impact count functions $N_{\text{imp}}^{(T_e>P)}$, panel (d), exhibit distinct maxima: $^4$Ca around $t = $ -30 fs, $^{40}$Ca at the laser pulse peak $t = 0$ and $^{131}$Ca with a considerable delay and broader maximum around $t = 100$ fs. With increasing Ca mass, the ionization of the first He atom is shifted to later times as histograms of the $t_{\text{He-1}}$ times in panel (e) show. Without the positive feedback of the He ionization on the nanoplasma electron population (He ionizations being disabled in the MD simulations), the $N_{\text{imp}}^{(T_e>P)}$ functions are primarily an indicator for the opening of the EII channel of He. But since an open EII channel is also a necessary condition for the EII avalanche, one may try to relate the total number of impacts $N_{\text{imp}}^{(T_e>P)}$ along trajectories and averaged over the trajectory set as a measure for the ignition propensity. Thus, we define the time-integrated impact count

$$Z_{T_e>P} = \int_{-\infty}^{+\infty} N_{\text{imp}}^{(T_e>P)}(t) dt \qquad (2)$$

as a measure for the ignition propensity of the doped He droplet. The total impact counts $Z_{T_e>P}$ decrease markedly in the order $^{131}$Ca > $^{40}$Ca > $^4$Ca, 276, 176 and 50, respectively (not shown), implying that the ignition capability decreases with decreasing dopant mass. This conclusion is indeed confirmed by MD simulations where He ionization is enabled; the corresponding ignition probabilities in this example are 1, 0.45 and 0, respectively. Thus, taking up the discussion of the efficiency of different dopant species to ignite a He nanoplasma, we shall add to the list of crucial properties the dopants' atomic mass.

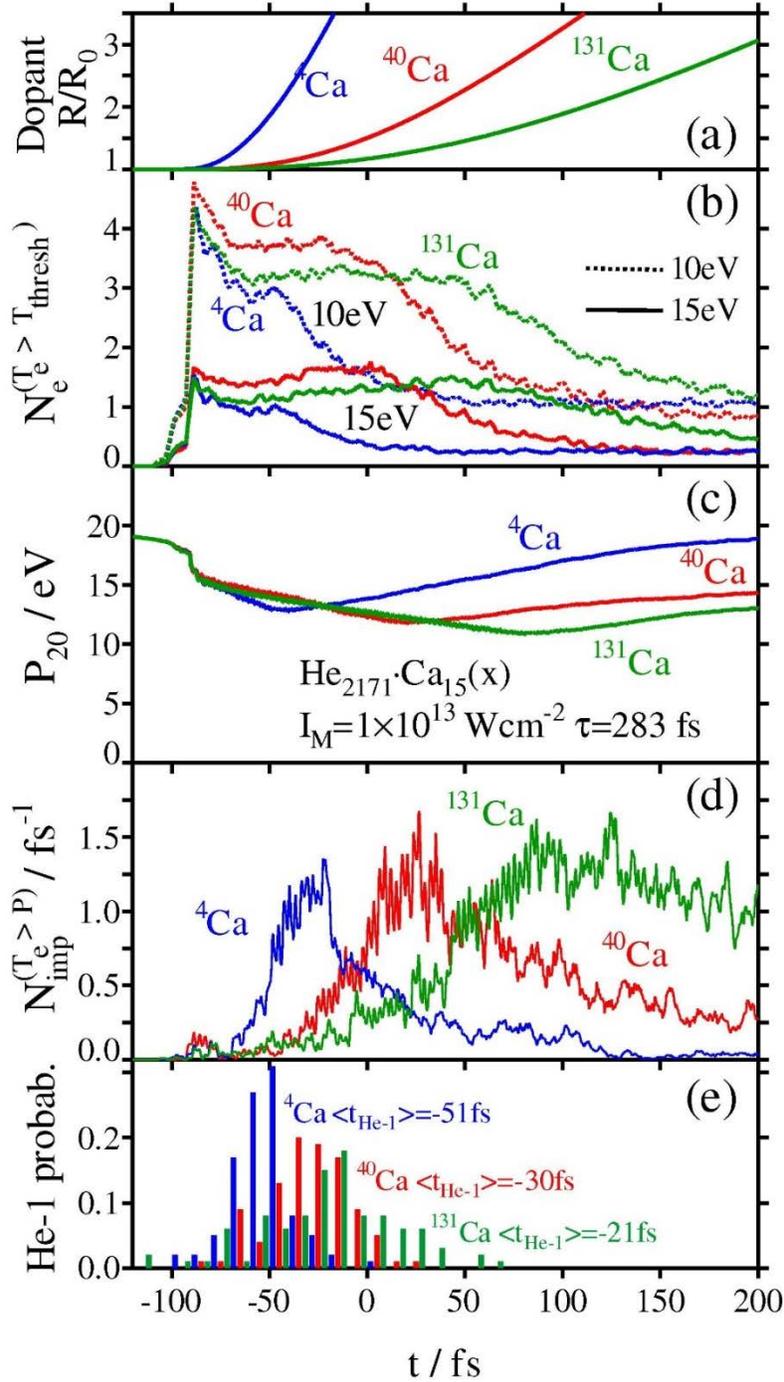

Figure 6: The effect of the dopant mass on the dopant expansion and several time-dependent properties characterizing the capability to open the EII channel of He, demonstrated for $He_{2171} \cdot Ca_{15}(x)$ at $I_M = 1 \times 10^{13}$ Wcm$^{-2}$. (a) The relative dopant cluster radius $R/R_0$, (b) the electron populations

$N_\text{e}^{(T_\text{e}>T_\text{thresh})}$ above threshold energies $T_\text{thresh}$ of 10 and 15 eV, (c) the ionization energies $P_{20}$ of the ionization energy distributions of neutral He atoms, (d) the impact count functions $N_\text{imp}^{(T_\text{e}>P)}$ and (e) the distributions of times $t_\text{He-1}$ with average trajectory-set average values $\langle t_\text{He-1} \rangle$ indicated in the panel. Results are presented for real Ca masses (red curves) and fictitious masses of 4 (blue) and 131 (green) amu. For the results in panels (a)-(d), He ionization was disabled to assess the dopant properties alone and the results were averaged over 50 trajectories. In addition, the functions $N_\text{e}^{(T_\text{e}>T_\text{thresh})}$ and $N_\text{imp}^{(T_\text{e}>P)}$ were short-time averaged over 4 fs to smooth oscillations. The distributions of $t_\text{He-1}$ times, panel (e), were obtained from sets of 100 trajectories of regular MD simulations in which He ionization was enabled.

*3.4 Comparison of Ca and Xe dopants*

Xe is characterized by a higher first ionization energy of 12.1 eV compared to 6.1 eV for Ca, a higher atomic mass (131 compared to 40 amu for Ca), the occupation of interior doping sites in the droplet and a by shorter Xe-He interatomic distance of 4.15 Å compared to a distance Ca-He of 5.9 Å. To assess the dopant properties of Xe and to compare them with Ca, in figure 7 we show the dopant cluster expansion $R/R_0$, the $P_{20}$ time evolution, the electron populations above 10 and 15 eV, and the $N_\text{imp}^{(T_\text{e}>P)}$ counts, for He$_{2171}$·Xe$_8$(c) (green curves), He$_{2171}$·Ca$_8$(x) (red curves) and He$_{2171}$·Ca$_8$(x) with an artificially delayed Ca ionization (blue curves) in panels (b) – (e), analogously to figure 6(a-d). In addition, we have included the average dopant charge per atom in panel (a). The higher laser pulse peak intensity of $I_\text{M} = 1 \times 10^{14}$ Wcm$^{-2}$ is chosen here because of the higher ionization energy of Xe, requiring intensities of $I_\text{M} \gtrsim 5 \times 10^{13}$ Wcm$^{-2}$ for TI. As in figure 6, He ionization is disabled to assess exclusively the effect of the dopants.

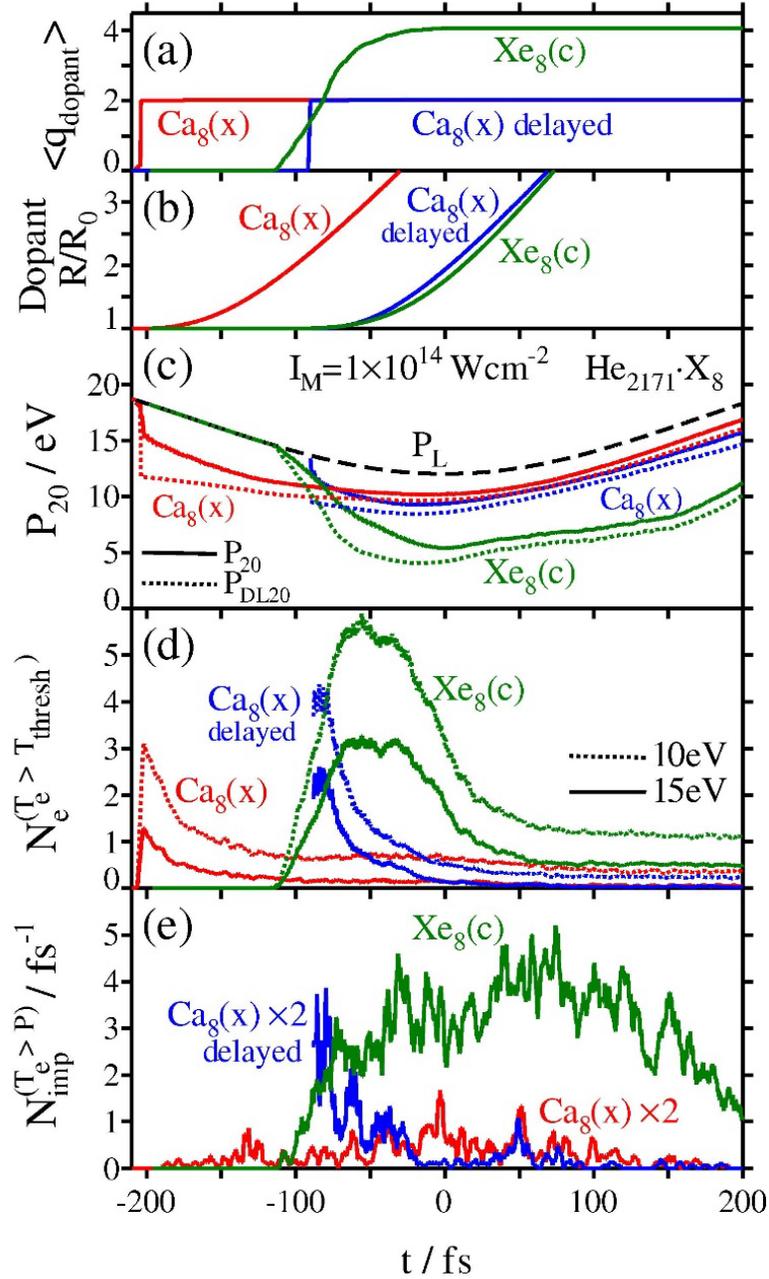

Figure 7: Comparison of the ignition capabilities of Xe and Ca for $He_{2171} \cdot Xe_8(c)$ (green curves) and $He_{2171} \cdot Ca_8(x)$ (red curves) at $I_M = 10^{14}$ Wcm$^{-2}$. Results for $He_{2171} \cdot Ca_8(x)$ where the Ca ionization is artificially delayed to the instant of the Xe ionization ($t_{Xe-1} = -91$ fs) are included to show the effect of the incubation time on the ignition capability (blue curves). (a) The average dopant charge per atom. (b) The relative dopant radii $R/R_0$. (c) The $P_{20}$ He ionization energies (closed lines), the $P_{DL20}$ He ionization energies (dotted lines) where only the electric field of the dopant cations and of the laser determine the Coulomb barrier at neutral He, and the ionization energies $P_L$ (black dashed line) where only the laser electric field is considered. (d) Electron populations $N_e^{(T_e > T_{thresh})}$ with kinetic energies above $T_{thresh} = 10$ and $15$ eV. (e) Impact counts $N_{imp}^{(T_e > P)}$. The values for He$_{2171}$·Ca$_8$(x) are multiplied by 2 to improve visual clarity. He ionization was disabled to assess the dopant properties alone. All results are averaged over 50 trajectories.

As the average dopant charges in figure 7(a) show, Xe is ionized later than Ca during the laser pulse because of its higher ionization energy but then donates four seed electrons per atom compared to two per Ca atom. This can be rationalized by the lower third and fourth ionization energy of Xe (32.1 and 46.7 eV compared to 50.9 and 67.1 eV for Ca). The higher Xe charges lead to a drastic downshifting of $P_{20}$ below the level $P_L$ (dashed black line) which would be achieved by the external laser electric field alone and also far below the level reached by the Ca$_8$ dopant (panel c). To indicate the effect of the shielding of the dopant cation charges by the seed electrons, also the $P_{DL20}$ functions are included which are obtained only by the sum of laser and dopant cation electric fields (dotted curves in panel c). Thus, when the shielding effect of electrons is disregarded, ionization energies are further downshifted. However, the net lowering of $P_{20}$ for Xe$_8$ including the shielding effect still largely exceeds that of the completely deshielded Ca$_8$ dopants.

The Ca$_8$ and the Xe$_8$ dopant clusters expand spatially at a very similar rate (figure 7(b)). This becomes particularly visible when comparing the $R/R_0$ ratio of Xe$_8$ (green curve) with the one of Ca$_8$ where the Ca ionization is artificially delayed (blue curve) to the instant ($t$ = -91 fs) at which Xe is ionized. In the cluster vertical ionization approximation, the time to double the cluster radius is $\propto \sqrt{m}/q$ [27] where m and q are the atomic mass and charge, respectively. With average atomic charges of 2 and 4, the $\sqrt{m}/q$ ratios for Ca and Xe are similar, 3.2 and 2.9, respectively. Thus, although in principle the larger mass favors Xe, it just compensates for the effect of the higher Xe charge on the dopant cluster expansion rate.

The nanoplasma electron populations with kinetic energies above 10 and 15 eV, figure 7(d), are at their peaks around twice higher for Xe than for Ca. Moreover, the peaks for Ca are close to the instants of Ca ionization at an early stage of the laser pulse; the populations decay rapidly and are small when the He ionization energies are downshifted by deshielding. As a consequence of the electron populations, the lowering of the He ionization energy and the interior doping site, the impact counts $N_{\text{imp}}^{(T_e>P)}$ in panel (e) are much higher for Xe than for Ca. When the Ca ionization is artificially delayed to the instant at which Xe is ionized on average $\langle t_{\text{Xe-1}} \rangle$ = -91 fs), $P_{20}$ decreases more rapidly and the impact counts considerably increase (blue curves) but are still below the values of Xe doping.

The much higher capability of Xe to open the EII channel of He also becomes apparent when looking at the trajectory set-averaged and time-integrated impact counts $Z_{T_e>P}$, Eq. (2). In table 1 the $Z_{T_e>P}$ values are listed together with the minimum trajectory set-averaged $P_{20}$ values and ignition probabilities for Xe$_8$, Ca$_8$ and other Ca dopant clusters discussed in the previous sections. From table 1 we infer:

(i) Xe$_8$ has much higher $Z_{T_e>P}$ counts than Ca$_8$ and even competes at $I_M = 10^{14}$ Wcm$^{-2}$ with the larger Ca$_{15}$ dopant, even when it is brought to the droplet surface (x, y) and at the longer He-Ca distance of 5.9 Å. Even in surface location and at the He-Xe distance of 5.9 Å, the ignition probability stays close to 1.

(ii) The trend of the $Z_{T_e>P}$ counts grossly follows the ignition probabilities. In most examples 200-300 $Z_{T_e>P}$ counts are sufficient to achieve an ignition probability of 1. The $Z_{T_e>P}$ counts increase with increasing the dopant cluster size and reproduce the trend C > X > Y of the ignition probabilities.

(iii) However, there are also marked deviations from a correlation between the $Z_{T_e>P}$ values and ignition probabilities: Not in all cases the $Z_{T_e>P}$ values increase with increasing pulse peak intensity as the ignition probabilities do. Although the artificial delay of the Ca$_8$ ionization leads to a more pronounced $N_{\text{imp}}^{(T_e>P)}$ peak, it does not yield higher $Z_{T_e>P}$ counts. The latter indicates the limitation of relating $Z_{T_e>P}$ to the ignition probability and that the temporal profile of $N_{\text{imp}}^{(T_e>P)}$ with respect to the pulse peak matters.

(iv) The minima of the trajectory set-averaged $P_{20}$ functions are lower for x than for y orientation (for Ca doping by a few tenths of an eV, for Xe even by up to 2 eV).

Table 1: Impact counts $Z_{T_e>P}$, minimum He ionization energies $P_{20}$ and ignition probabilities of selected doped He droplets.

| System | $I_M$/Wcm$^{-2}$ | $Z_{T_e>P}$ [a] | Min. $P_{20}$/eV [a] | Ignition Prob. [b] |
|---|---|---|---|---|
| He$_{2171}$·Ca$_8$(c) | 1×10$^{14}$ | 149 | 9.7 | 0.70 |
| He$_{2171}$·Ca$_8$(x) | 1×10$^{14}$ | 83 | 10.2 | 0.27 |
| He$_{2171}$·Ca$_8$(y) | 1×10$^{14}$ | 51 | 10.6 | 0.25 |
| He$_{2171}$·Ca$_8$(x) delayed [c] | 1×10$^{14}$ | 70 | 9.3 | 0.64 |
| He$_{2171}$·Ca$_{15}$(x) | 1×10$^{13}$ | 136 | 11.7 | 0.45 |
| He$_{2171}$·Ca$_{15}$(x) | 2×10$^{13}$ | 162 | 11.0 | 0.91 |
| He$_{2171}$·Ca$_{15}$(x) | 5×10$^{13}$ | 193 | 10.2 | 0.97 |
| He$_{2171}$·Ca$_{15}$(x) | 1×10$^{14}$ | 240 | 9.4 | 1 |
| He$_{2171}$·$^4$Ca$_{15}$(x) [d] | 1×10$^{13}$ | 50 | 12.9 | 0 |
| He$_{2171}$·$^{131}$Ca$_{15}$(x) [d] | 1×10$^{13}$ | 276 | 11.5 | 0.46 |
| He$_{2171}$·Ca$_{15}$(y) | 1×10$^{13}$ | 126 | 12.2 | 0.31 |
| He$_{2171}$·Ca$_{15}$(y) | 2×10$^{13}$ | 139 | 11.1 | 0.71 |
| He$_{2171}$·Ca$_{15}$(y) | 5×10$^{13}$ | 144 | 10.4 | 0.88 |
| He$_{2171}$·Ca$_{15}$(y) | 1×10$^{14}$ | 161 | 10.0 | 0.98 |
| He$_{2171}$·Ca$_{23}$(x) | 1×10$^{13}$ | 630 | 9.9 | 0.99 |
| He$_{2171}$·Ca$_{23}$(x) | 2×10$^{13}$ | 635 | 8.8 | 1 |
| He$_{2171}$·Ca$_{23}$(x) | 5×10$^{13}$ | 639 | 7.8 | 1 |
| He$_{2171}$·Ca$_{23}$(x) | 1×10$^{14}$ | 620 | 7.3 | 1 |
| He$_{2171}$·$^4$Ca$_{23}$(x) [d] | 1×10$^{13}$ | 158 | 11.8 | 0.09 |
| He$_{2171}$·$^{131}$Ca$_{23}$(x) [d] | 1×10$^{13}$ | 1226 | 9.6 | 1 |
| He$_{2171}$·Ca$_{23}$(y) | 1×10$^{13}$ | 573 | 10.7 | 0.99 |
| He$_{2171}$·Ca$_{23}$(y) | 2×10$^{13}$ | 511 | 10.0 | 1 |
| He$_{2171}$·Ca$_{23}$(y) | 5×10$^{13}$ | 477 | 9.1 | 1 |
| He$_{2171}$·Ca$_{23}$(y) | 1×10$^{14}$ | 436 | 8.5 | 1 |
| He$_{2171}$·Xe$_8$(c) | 1×10$^{14}$ | 1301 | 5.3 | 1 |
| He$_{2171}$·Xe$_8$(x, 4.15Å) [e] | 1×10$^{14}$ | 423 | 7.7 | 1 |
| He$_{2171}$·Xe$_8$(x, 5.9Å) [e] | 1×10$^{14}$ | 370 | 8.0 | 1 |
| He$_{2171}$·Xe$_8$(y, 4.15Å) [e] | 1×10$^{14}$ | 211 | 9.6 | 0.98 |
| He$_{2171}$·Xe$_8$(y, 5.9Å) [e] | 1×10$^{14}$ | 179 | 10.1 | 0.87 |

(a) $Z_{T_e>P}$ and minimum $P_{20}$ values were obtained from averages over 50 trajectories with disabled He ionization to assess the effect of the dopants alone, (b) the ignition probabilities from sets of 100 trajectories, (c) Ca ionization artificially delayed to the instant of Xe ionization (section 3.4), (d) results for fictitious Ca masses 4 and 131 discussed in section 3.3. (e) Xe$_8$ placed at the droplet surface parallel (x) and perpendicular (y) to the laser polarization with Xe-He distances of 4.15 and 5.9 Å, respectively.

In summary, there are two main reasons for the higher ignition efficiency of Xe: the larger number of seed electrons per atom (four compared to two per Ca atom) and the much larger and much more rapid downshift of the Coulomb barriers by the electric field of the higher Xe cation charges. The latter effects that the population of energetic seed electrons is still high when the Coulomb barriers at He are sufficiently suppressed. Furthermore, the interior doping site and the shorter He-Xe distance are favorable factors, as discussed previously [13]. However, we find that they are of minor importance in this particular example.

By carrying out comparative MD simulations without Pauli repulsion we have also assessed its effect on the ignition probability. We found for Ca dopants that it requires in many cases one Ca atom less to achieve the same ignition probability as with Pauli repulsion. When the Pauli potential was switched on, Ca$_8$(c) in interior doping position (not shown in figure 7) causes a somewhat slower spatial expansion than Ca$_8$(x) in surface position. We attribute this effect to somewhat stronger shielding of the Ca charges by the seed electrons in Ca$_8$(c). Consistently, in Ca$_8$(c), $P_{20}$ assumes slightly higher values at short times and slightly lower values at longer times, the latter because of the slower expansion of the Ca$_8$(c) dopant cluster. Presumably these small differences originate from the Pauli repulsive potential on He which confines the seed electrons for interior doping to some extent.

*3.5 The onset of avalanche ionization*

We note that an EII avalanche He$^{q+}$ + e → He$^{(q+1)+}$ + 2e resembles an autocatalytic reaction A + B → C +2B, i. e., a reaction in which one of the products (B) is also a reactant, so that the course of the reaction has a positive feedback on its rate. In the EII avalanche, the electrons take the role of reactant B. He$^{q+}$ corresponds to reactant A and He$^{(q+1)+}$ to product C with the slight complication that the simple A + B → C + 2B reaction is replaced by a two-step reaction, He + e → He$^+$ + 2e and He$^+$ + e → He$^{2+}$ + 2e. The number $N_{ii}$ of inner ionizations or the average He charge $\langle q_{He} \rangle$ corresponds to the amount of C. Characteristic for an autocatalytic reaction is the sigmoidal shape of the amount or concentration of C, with a slow rising edge termed *induction period*. Indeed, the simulated $\langle q_{He} \rangle$ curves are sigmoidal (see, for example, figure 1(c)) in ref. [14]), with the induction period resembling the incubation time. In principle, one could attempt to map the MD simulation results for the time evolution of the number $N_{He^{q+}}$ of He atoms in their charge states $0 \leq q \leq 2$ and the number $N_e$ of nanoplasma electrons on a set of rate equations. This implies a subdivision of the incubation time into two parts, a period $t_{\text{He-1}}$ − $t_{\text{dopant-1}}$ during which the EII channel of He is closed and the rate constants are zero, and a second part $t_{\text{ignit}} - t_{\text{He-1}}$ when the rate equations describe the evolution of the EII avalanche. We refrain here from a detailed analysis in terms of the rate equations and content ourselves with establishing the time scales and a few qualitative remarks.

For several pulse peak intensities in the range $10^{13} - 10^{14}$ Wcm$^{-2}$, figure 8 exhibits the trajectory set-averaged $\langle t_{\text{He-1}} - t_{\text{dopant-1}} \rangle$ times in panel (a), the $\langle t_{\text{ignit}} - t_{\text{He-1}} \rangle$ times in panel (b) and the underlying $\langle t_{\text{dopant-1}} \rangle$, $\langle t_{\text{He-1}} \rangle$ and $\langle t_{\text{ignit}} \rangle$ times for Ca and Xe in panels (c) and (d) as a function of the dopant cluster size, with the $\langle t_{\text{ignit}} \rangle$ times based on the $\langle q_{\text{He}} \rangle = 0.1$ criterion for ignition. The results exhibit the following features:

(i) As discussed in section 3.4, the $\langle t_{\text{He-1}} - t_{\text{dopant-1}} \rangle$ times, panel (a), are much longer for Ca (40-130 fs) than for Xe (<10 fs), reflecting the slow deshielding process of the Ca cation charges and lesser downshifting of He ionization energies.

(ii) For constant pulse peak intensity, the $\langle t_{\text{He-1}} - t_{\text{Ca-1}} \rangle$ times are highly dopant cluster size dependent as a result of a strong cluster size dependence of the $\langle t_{\text{He-1}} \rangle$ and a near cluster size independence of the $\langle t_{\text{Ca-1}} \rangle$ times (panel (c)). The earlier $\langle t_{\text{He-1}} \rangle$ times for larger dopant clusters are caused by a larger and more rapid downshift of the ionization potential of He. The near independence of the $\langle t_{\text{Ca-1}} \rangle$ times of the dopant cluster size is a consequence of the small TI probabilities of Ca. A single TI in the dopant cluster is sufficient to trigger a nearly instantaneous ionization of the entire dopant cluster. Since the cumulative probability of a single TI increases with the number of dopant atoms, $\langle t_{\text{dopant-1}} \rangle$ is expected to advance to earlier times with increasing the dopant cluster size. However, the effect is small for Ca because of its low TI probability. In contrast, for the higher TI probability of Xe a marked advancement of $\langle t_{\text{Xe-1}} \rangle$ does take place. Once the Xe dopant cluster is ionized, the opening of the EII channel of He follows within femtoseconds, so that the $\langle t_{\text{He-1}} - t_{\text{Xe-1}} \rangle$ times are short and nearly constant over the entire considered dopant cluster size range.

(iii) The $\langle t_{\text{ignit}} - t_{\text{He-1}} \rangle$ times, panel (b), are much shorter for Xe than for Ca doped droplets. We attribute this result to the larger EII cross sections of He caused by the larger downshifting of the He ionization and to the closer proximity of the $\langle t_{\text{ignit}} - t_{\text{He-1}} \rangle$ periods to the pulse peak maximum, both causing higher He ionization rates.

(iv) The $\langle t_{\text{ignit}} - t_{\text{He-1}} \rangle$ times, panel (b), for both Ca and Xe shorten with increasing the dopant cluster size and with increasing the pulse peak intensity, mainly because of a shift of the $\langle t_{\text{ignit}} \rangle$ times (panels (c) and (d)). For Ca dopants the $\langle t_{\text{ignit}} - t_{\text{He-1}} \rangle$ times constitute around 50% of the total incubation time, for Xe dopants 80% and more.

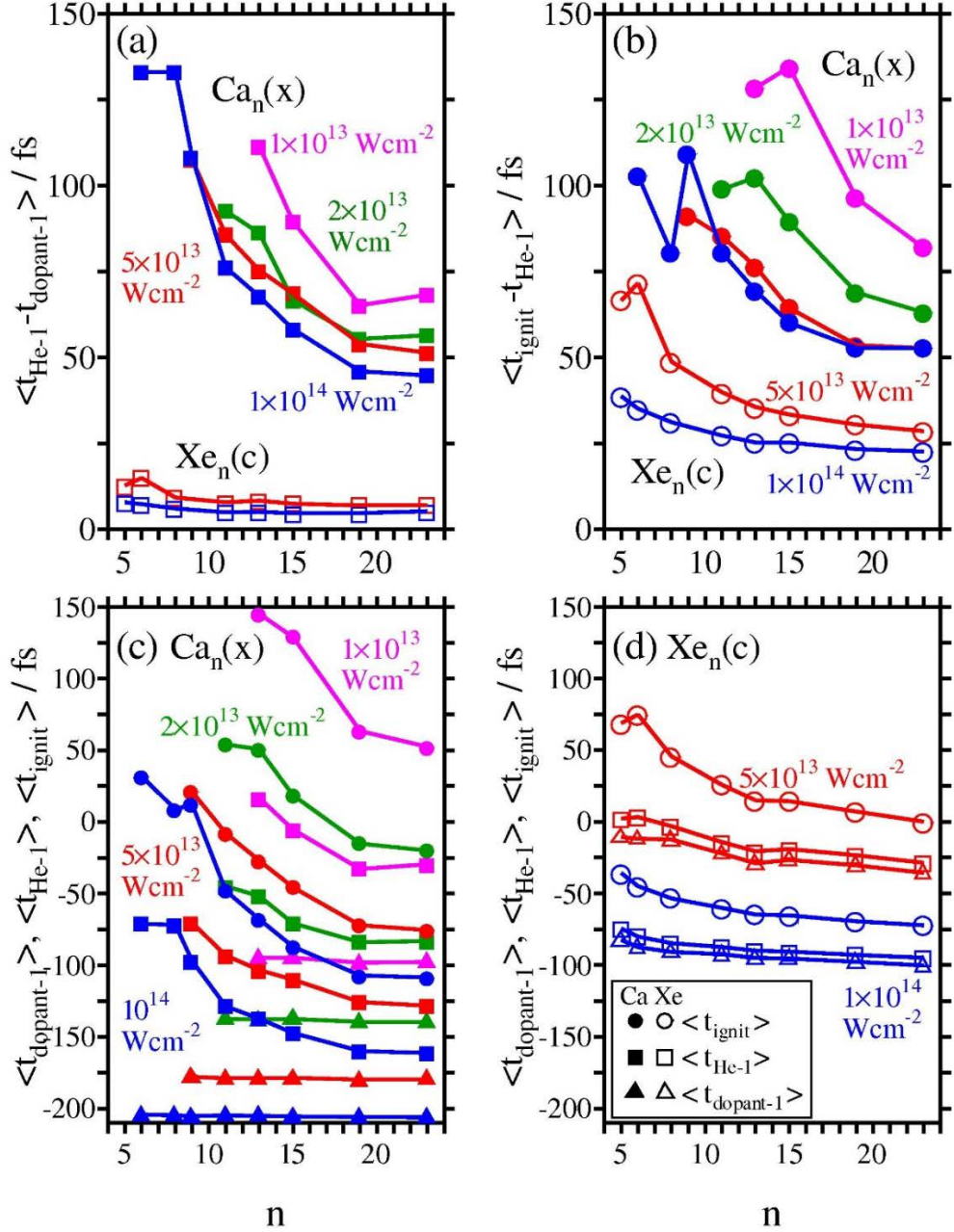

Figure 8: Dopant cluster size dependence of (a) the average times $\langle t_{\text{He-1}} - t_{\text{dopant-1}} \rangle$ between the dopant ionization and the ionization of the first He atom and (b) the average times $\langle t_{\text{ignit}} - t_{\text{He-1}} \rangle$ between the ionization of the first He atom and the instant of ignition for $He_{2171}$ droplets doped with $Ca_n$ (surface states x) and $Xe_n$ (interior states) clusters. The underlying average instants $\langle t_{\text{dopant-1}} \rangle$, $\langle t_{\text{He-1}} \rangle$ and $\langle t_{\text{ignit}} \rangle$ are represented in panels (c) for $Ca_n$ (closed symbols) and in panel (d) for $Xe_n$ dopant clusters (open symbols). The $\langle t_{\text{ignit}} \rangle$ times are based on the $\langle q_{\text{He}} \rangle = 0.1$ threshold criterion.

It should be noted that $t_{\text{ignit}}$, as it is defined here by the $\langle q_{\text{He}} \rangle = 0.1$ criterion, represents an early stage of the EII avalanche but not its very beginning. In our previous work [14] we showed that for every single trajectory, there is a close correlation between the instant of ignition $t_{\text{ignit}}$ (based on the $\langle q_{\text{He}} \rangle = 0.1$ criterion being equivalent to $N_{\text{He}^+} \approx 200$ He ionizations for He$_{2171}$ droplets considered in this work) and the final average He charge $\langle q_{\text{He, final}} \rangle$ at the end of the laser pulse, irrespective of the dopant element, dopant cluster size and dopant location in the droplet. The correlation is a manifestation of the ionization avalanche: duration and intensity profile of the laser pulse alone driving the chain reaction determine the product charge state composition. In an attempt to determine the very beginning of the avalanche, one may ask what is the minimum number $N_{\text{He}^+}$ of ionized He atoms needed for such a correlation $\langle q_{\text{He, final}} \rangle$ vs. $t_{N_{\text{He}^+}}$ to exist. We found that for $I_M \geq 5 \times 10^{13}$ Wcm$^{-2}$ the correlation holds down to $N_{\text{He}^+} = 40$ He atoms. When decreasing $N_{He^+}$ further, the correlation deteriorates gradually: For $N_{\text{He}^+} = 12$, still most of the trajectories obey the correlation and for $N_{\text{He}^+} < 10$ the correlation deteriorates massively. For the lower pulse peak intensities $I_M = 2 \times 10^{13}$ and $10^{13}$ Wcm$^{-2}$ (achieved here only for Ca doping) the lower limits of $N_{\text{He}^+}$ where the $\langle q_{\text{He, final}} \rangle$ vs. $t_{N_{\text{He}^+}}$ holds perfectly, can be drawn at $\approx 50$ and 80-100 He atoms, respectively. Particularly at $I_M = 10^{13}$ Wcm$^{-2}$ the He ionizations set in so late in many trajectories that it was difficult for us to recognize whether an avalanche ionization was terminated by the fading pulse or whether the avalanche has not started. Under such conditions ignition becomes a fuzzy concept. However, irrespective of the shortcomings to define the very beginning of the EII avalanche unambiguously, the $t_{\text{ignit}}$ times derived from the $\langle q_{\text{He}} \rangle = 0.1$ criterion are useful to compare trends among dopant elements, dopant cluster sizes and pulse peak intensities.

## 4. Conclusions

In this theoretical study we have addressed various mechanistic aspects of the initiation of EII avalanches (*ignition*) in He nanodroplets doped with small Ca and Xe clusters. The incubation time, which is the time $t_{\text{ignit}}$-$t_{\text{dopant-1}}$ between the initial dopant ionization and ignition, can be subdivided into two parts, (A) the time $t_{\text{He-1}}$-$t_{\text{dopant-1}}$ between the dopant ionization until the opening of the EII channel of He, marked by the ionization of the first He atom, and (B) the subsequent time $t_{\text{ignit}}$-$t_{\text{He-1}}$ to evolve into an EII avalanche. We have assessed the capability to open the EII channel of He and to ignite the He nanoplasma by defining an impact count function which quantifies the spatial overlap of the trajectories of the impinging seed electrons with neutral He atoms and the energetic overlap of the electron kinetic energy distribution with the distribution of He ionization energies.

The analysis in this paper addresses mainly the first part of the incubation time $t_{\text{He-1}}$-$t_{\text{dopant-1}}$, i. e., the opening of the EII channel of He. We have identified two effects which crucially assist the initial ionization of He: (i) The *deshielding* of the dopant cations: a partial loss of the seed electrons in the dopant-droplet interface region which reduces the shielding of the dopant cation charges and therefore increases the electric field particularly at the neighboring He atoms, thereby reducing their Coulomb barriers further. (ii) A *slingshot effect* by the dopant cations: the acceleration of seed electrons in the potential well of the dopant cations which assists their laser energy absorption drastically, so that seed electrons attain kinetic energies far above the ponderomotive energies of free electrons. This provides sufficient impact energies for EII and kicks part of the seed electrons far away from the active dopant-droplet interface region.

The spatial expansion of the dopant cluster weakens its electric field and therefore its capability to lower the Coulomb barriers of He, as well as its ability to act as a slingshot for electrons. Consequently, the dopant's ignition capability depends also on the atomic mass, as shown by our simulations. However, in the case of Ca and Xe, the higher Xe mass just cancels out the effect of the higher Xe ion charge on the expansion rate of the dopant; Ca and Xe dopant clusters expand by nearly the same rate.

We have identified two main reasons for the higher ignition efficiency of Xe: the larger number of seed electrons per dopant atom (four compared to two per Ca atom) and the much larger and much more rapid downshift of the Coulomb barriers of He by the electric field of the higher Xe cation charges. Due to the rapid downshift, the populations of energetic seed electrons are still high when the Coulomb barriers at He are sufficiently suppressed. Less important factors are the interior doping site and the shorter He-Xe distance, at least for the examples studied in this work.

In particular in view of designing highly efficient dopant systems with tailored properties it is desirable to characterize the ignition capabilities and the time-dependent ignition behavior of dopant clusters by a few simple quantities that contain the essence of the underlying physical process. These quantities, such as the ignition capabilities themselves, will depend on static dopant properties (electronic properties, atomic mass, dopant location) as well as laser pulse parameters; static dopant properties and pulse parameters cannot be decoupled, since for example deshielding of dopant charges and the dopant cluster expansion rate also depend on the laser pulse parameters. The impact count functions and their time integrals proposed in this paper go into this direction of linking static dopant properties and laser pulse parameters to a non-static parameter characterizing the ignition capability. However, once the He ionization is initiated, the generated He ions and electrons take active part and cause a positive feedback in the course of the ionization process. To assess the ignition capabilities of the subsystem dopant + laser pulse alone, we therefore disabled the He ionization when evaluating the impact count function. The drawback of this approach is that the impact count functions in their current formulation primarily assess the dopants' capability to induce the ionization of the first He atom.

Consequently, the second part of the incubation time, the evolution into an EII avalanche during the period $t_{ignit}$-$t_{He-1}$, still lacks a rigorous analysis. It is clear that the number of energetic seed electrons which survived the first part of the incubation time ($t_{He-1}$-$t_{dopant-1}$) inside or close to the nuclear framework, determine the length of the second part of the incubation time, $t_{ignit}$-$t_{He-1}$. If the population of energetic seed electrons is massively depleted, ignition does not take place at all. This also demonstrates the ambivalence of the loss of seed electrons: On the one hand, the EII rate increases with the number of seed electrons. On the other hand, a loss of seed electrons promotes EII by reducing the Coulomb barriers of He. Both objectives are met simultaneously only, if the dopant cluster is large enough. Since the nature of the EII avalanche is a positive feedback of the He ionization on the ionization chain reaction, whereas the impact count function in its current formulation does not take into account He ionizations, the impact count function is limited to providing a measure for the starting conditions of a subsequent avalanche, but cannot predict the result of the avalanche itself. For this reason, the impact count function values only grossly indicate an ignition propensity but cannot predict ignition probabilities accurately. A possible way to go one step beyond could be solving a set of rate equations for the populations of He, $He^+$, $He^{2+}$ and nanoplasma electrons, even if the rate constants will show some time dependence (in particular because the He ionization energies will vary with the He charge state abundances and other nanoplasma properties).

**Acknowledgments**

The authors thank for financial support from the Spanish Ministerio de Economia y Competivitad (ref. no. CTQ2015-67660-P) and from the Deutsche Forschungsgemeinschaft (DFG) within the project MU 2347/12-1 in the frame of the Priority Programme 1840 'Quantum Dynamics in Tailored Intense Fields'. Computational and manpower support provided by IZO-SGI SG Iker of UPV/EHU and European funding (EDRF and ESF) is gratefully acknowledged.